\documentstyle[11pt,moriond,epsfig]{article}

\def\be{\begin{equation}}
\def\ee{\end{equation}}
\def\bea{\begin{eqnarray}}
\def\eea{\end{eqnarray}}
\def\nn{\nonumber}

\def\lsim{\raise0.3ex\hbox{$<$\kern-0.75em\raise-1.1ex\hbox{$\sim$}}}
\begin{document}
\vspace*{4cm}
\title{Charged Multiplicities at SPS and RHIC and 
consequences for $J/\psi$ suppression}

\author{ A. CAPELLA, {\underline {D. SOUSA}} }

\address{Laboratoire de Physique Th\'eorique,\\
Universit\'e de Paris-Sud, B\^atiment 210, F-91405 Orsay Cedex, France}

\maketitle\abstracts{Hadron multiplicities in nucleus--nucleus 
interactions are calculated in the Dual Parton Model and 
its dependence on the number of collisions and the number of participants is analyzed.
Shadowing corrections are calculated as a function of impact parameter
and the multiplicity per participant as a funtion of centrality is 
found to be in 
agreement with experiment at SPS and RHIC energies. 
The obtained results are used to compute the $J/\psi$ suppression
in a comover approach.}


\section{Charged Multiplcities in the Dual Parton Model}

In the Dual Parton Model (DPM) the charged multiplicity per unit rapidity
in a symmetric collision is given by \cite{1r} 

\begin{eqnarray}
\label{1e}
{dN_{AA}^{ch} \over dy}(y,b) & = & n_A (b) \left [ N_{\mu}^{qq^{P}-q_v^T}(y) +
N_{\mu}^{q_v^P-qq^T}(y) + (2k - 2) N_{\mu}^{q_s-\bar{q}_s} \right ] + \nn \\ 
&&\Big ( n(b) - n_A(b)\Big ) \Big ( 2 k \ N_{\mu}^{q_s-\bar{q}_s}(y)
\Big ) \quad .
\end{eqnarray}

\noindent
Here $P$ and $T$ stand for the projectile and target nuclei, $n(b)$
is the average number of binary collisions and $n_A (b)$
the average number of participants of nucleus $A$. These quantities 
can be computed in a Glauber model. $k$ is
the average number of inelastic collisions in $pp$ and 
$\mu (b) = kn(b)/n_A(b)$ is the total average number of collisions suffered
by each nucleon.

In the DPM each inelastic collision leads to two strings and
the total number of strings is $2kn$. The charged multiplicity
produced by a single string is obtained by a convolution of momentum distribution
functions and fragmentation function, and it depends on $\mu$ due to energy
conservation. 
We see from (1) that the multiplicity is given by a linear combination
of $n_{A}$ and $n$ with coeficients which depend on the impact parameter
$b$ (via $\mu(b)$).

Shadowing corrections in Gribov theory are universal \cite{5r}, i.e. they apply both
to soft and hard processes. They are closely related to the size of diffractive
production and, thus, are controlled by triple Pomeron diagrams.
The reduction
of the multiplicity resulting from shadowing corrections has been computed in
\cite{5r}. These
corrections are negligeable at SPS energies
but at RHIC energies they reduce the multiplicity by 40 to 50\%.

\begin{figure}
\begin{minipage}{40mm}
\centering\epsfig{file=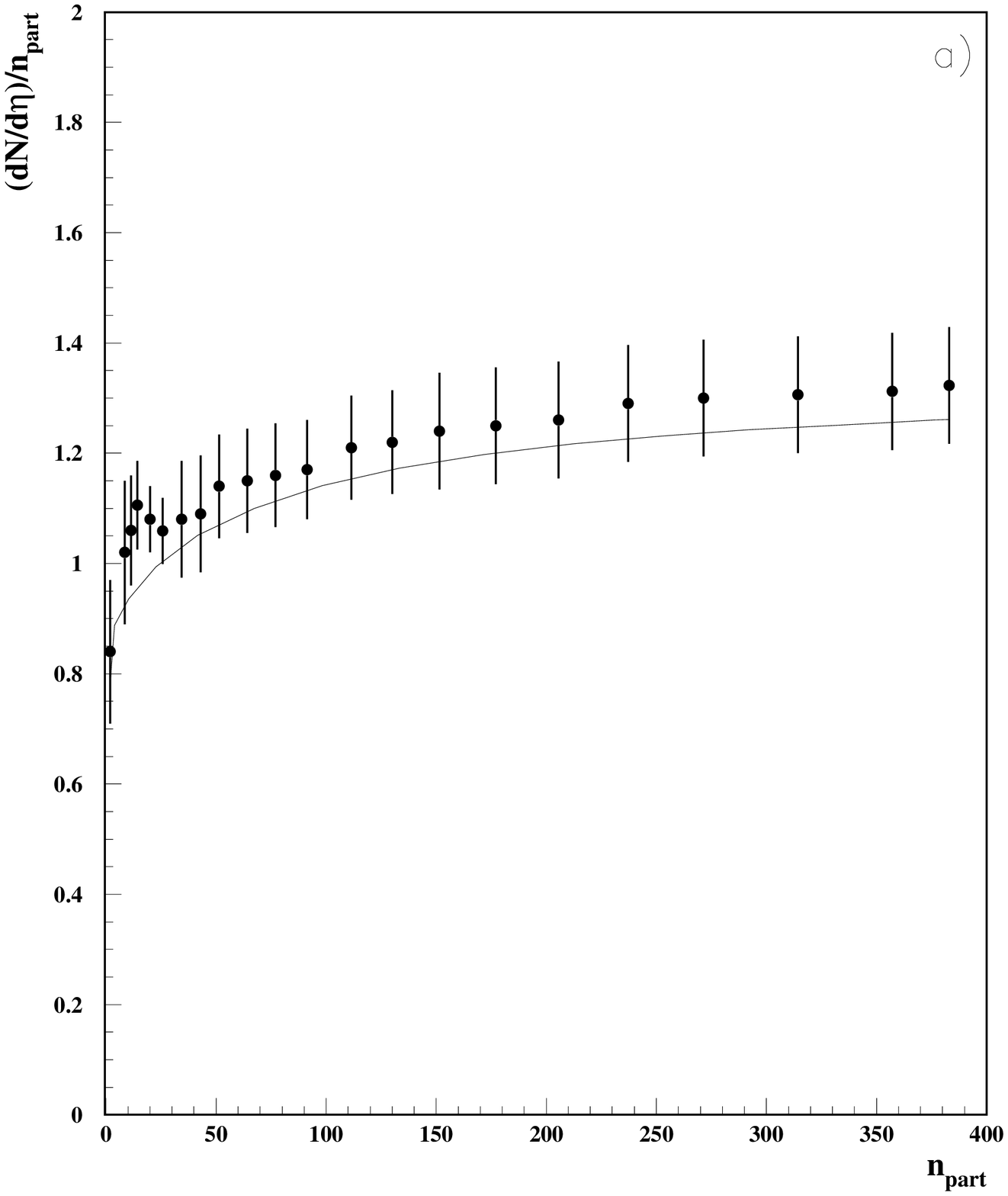,bbllx=0,bblly=30,bburx=480,bbury=620,height=3.0in}
\end{minipage}
\hspace{\fill}
\begin{minipage}{40mm}
\centering\epsfig{file=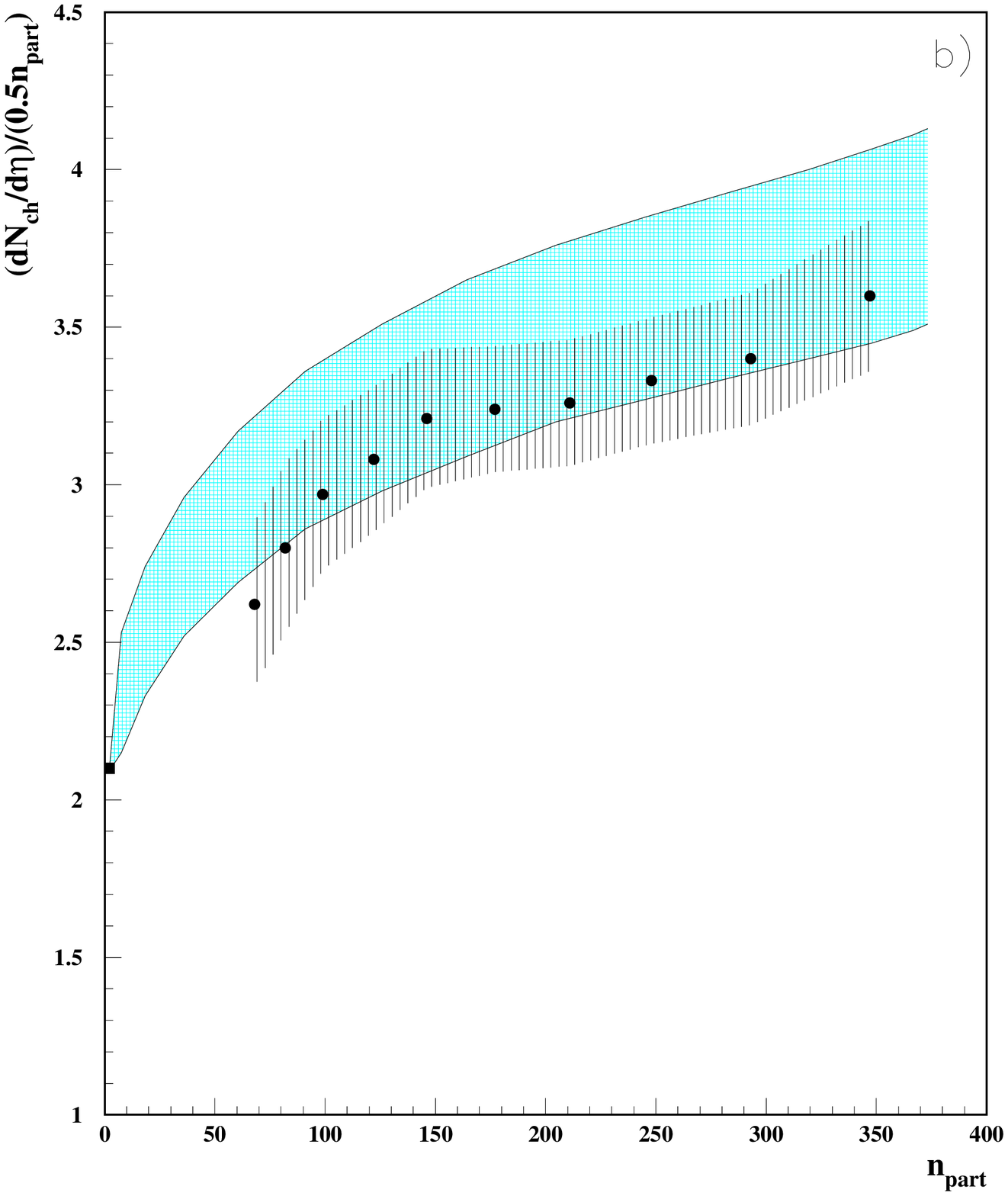,bbllx=280,bblly=30,bburx=840,bbury=620,height=3.0in}
\end{minipage}
\caption{
a): The values of $dN^{ch}/d \eta /n_{part}$
versus $n_{part}$ for
$PbPb$ collisions at $\sqrt{s}
= 17.3$~GeV in the range $- 0.5 < \eta_{cm} < 0.5$ computed from eq.
(\ref{1e}), compared with the 
WA98 data \protect\cite{10r}.
b): The values of $dN^{ch}/d \eta /(0.5 n_{part})$ for
$Au$-$Au$ collisions at
$\sqrt{s} = 130$~GeV in the range $- 0.35 < \eta_{cm} < 0.35$ computed
from eq. (\ref{1e}) taking into account shadowing corrections (see main
text). They are compared to the data \protect\cite{11r}.}

\end{figure}

We present the results \cite{1r} obtained at two
different energies~: $\sqrt{s} = 17.3$ and 130 GeV. The corresponding 
non-diffractive cross-sections are $\sigma_{ND} = 26$ and 33 mb,
respectively. We take $k = 1.4$ and 2.0 corresponding to $dN_{pp}^{ND}/dy = 1.56$
and 2.72. The result in absence of shadowing at $\sqrt{s} = 17.3$ is shown in Fig~1 (a) 
(solid line). We obtain a mild increase of the multiplicity per participant
consistent with the results of the WA98 Collaboration \cite{10r}.
This increase gets stronger with increasing energies. As we pointed out before,
shadowing corrections are negligeable at SPS energies but their effect 
is large at RHIC. Unfortunately, shadowing corrections have a rather large 
uncertainty at RHIC energies. Two alternative calculations \cite{1r} of
shadowing lead to the results at $\sqrt{s} = 130$ GeV 
shown by the solid lines in Fig~1 (b).
Clearly, with the larger values of the
shadowing corrections we obtain a quantitative agreement with the PHENIX data \cite{11r}.

Note that our calculations refer to $dN/dy$ while
the first RHIC measurements refer to $dN/d\eta$.
The latter is, of course, smaller at mid rapidities. This
difference is negligibly small as SPS where the laboratory
pseudo-rapidity variable is used. However, at $\sqrt{s} = 130$ GeV 
where $\eta_{cm}$ is used instead, their ratio can be as large as 1.3 \cite{13r}.

\section{$J/\psi$ suppression}

Now we are going to use these multiplicities, computed in the framework of DPM,
to determine the $J/\psi$ suppression in the comovers approach. 
In this model, the $J/\psi$ survival probability is the product of two factors 
$S_{abs}(b,s)\cdot S_{co}(b,s)$. The first factor represents the suppression due 
to nuclear absorption of the $c\bar{c}$ pair. Its expression, 
given by the probabilistic Glauber model, is
well known. It contains a parameter, the absorptive cross-section
$\sigma_{abs}$. The second factor $S_{co}(b,s)$ represents the suppression
resulting from the interaction with comovers. Its expression \cite{14r,16r}
depends on the averaged interaction cross-section $\sigma_{co}$ and 
on the value of the multiplicity at each $b$ in the region of the dimuon trigger
($0 < y^* < 1$), for which we take the results of section 1,
($N^{co}_{y_{DT}}$ is the charged multiplicity; the factor 3/2 takes care of
the neutrals)

\begin{equation}
\label{8e}
S_{co}(b,s) = \exp \left \{ - \sigma_{co} {3 \over 2} 
\ N^{co}_{y_{DT}}(b, s) \ \ln { \left [
{{3 \over 2} N^{co}_{y_{DT}}(b, s) \over N_{f}} \right ] } \right \} \ .
\end{equation}

\noindent
Experimentally, the ratio of $J/\psi$ over DY is plotted as a function
of either $E_T$ or the energy of the zero degree calorimeter $E_{ZDC}$. $E_T$
is the transverse energy of neutrals deposited in the NA50 calorimeter, located
in the backward hemisphere ($1.1 < y_{lab} < 2.3$). Using the proportionality
between $E_T$ and multiplicity, 
we have 

\begin{equation}
\label{9e}
E_T(b) = {1 \over 2} \ q \ N_{y_{cal}}^{co}(b) \ .
\end{equation}

\noindent
Here the multiplicity of comovers is
determined in the rapidity region of the NA50 calorimeter. 
The energy of the zero degree calorimeter is defined as

\begin{equation}
\label{10e} 
E_{ZDC}(b) = [A - n_A(b)] E_{in} + \alpha n_A(b) E_{in} \ , 
\end{equation}

Here $n_A$ is the
number of participants, $A - n_A$ the number of spectators, and
$E_{in}$ = 158~GeV is the beam energy. The last term represents the small
fraction of wounded nucleons and/or fast secondaries that hit the ZD Calorimeter.  
The values of $q$ and $\alpha$ are obtained, respectively, from a fit of
the tails of the $E_T$ and $E_{ZDC}$ distributions. With the obtained values,
the measured $E_T - E_{ZDC}$ correlation is well reproduced.  

We see from Eqs. (\ref{8e}) and
(\ref{9e}) that, in order to describe the centrality dependence of the $J/\psi$
suppression, it is paramount to have a good description of the $b$ dependence
of $N_y^{co}$ -- both in the rapidity region of the dimuon trigger and in the
one of the $E_T$ calorimeter. 
In the latter we obtain an scaling in the number of participants, i.e. we
recover the wounded nucleon model.

The model allows to compute the ratio $J/\psi$ over DY versus either $E_T$
or $E_{ZDC}$ from peripheral collisions up to the knee of the 
$E_T$ or $E_{ZDC}$
distributions. To go beyond it, we have to introduce \cite{16r,17r} the
fluctuations responsible for the tail of the $E_T$ and $E_{ZDC}$ distributions 
(Eqs. (3) and (4) give only the average values at each $b$). They have
been introduced \cite{16r,17r} in the model by multiplying
$N_{y_{DT}}^{co}(b)$ in Eq. (\ref{8e}) by $F(b) = E_T/E_T(b)$, where
$E_T$ is the measured value of the transverse energy. 
Likewise, for the suppression versus $E_{ZDC}$ we multiply $N^{co}_{y_{DT}}(b)$
by $n_{A}(E_{ZDC}) / n_{A}(b)$ where $n_{A}(E_{ZDC})$ is obtained
from Eq. (4): $n_{A}(E_{ZDC}) = (A E_{in} - E_{ZDC}) / E_{in} (1 - \alpha)$. 

The results are presented in Fig.~2 and compared with 
the NA50 data \cite{17br}. We
see that from peripheral collisions up to the knee of the $E_T$
distribution, the data are well described. However, beyond
the inflexion point at the knee, the decrease in the data is sharper
than in the model. Note, however, that the data beyond the knee are
obtained with the so-called minimum bias (MB) analysis. Only the ratio
$J/\psi$ over MB is measured and it is multiplied by a theoretical
ratio DY/MB. In the model used by NA50, this ratio 
is essentially flat beyond the knee -- due to the fact
that the tail of the $E_T$ distribution of hard (DY) and
soft (MB) processes is assumed to be the same. 
The behaviour of the MB NA50 data beyond the knee can be explained \cite{19r}
by the combined effects of a small decrease of the hadronic $E_T$ in
the $J/\psi$ event sample (due to the $E_T$ taken by the $J/\psi$ trigger),
together with the sharp decrease of the $E_T$ distributions in this tail
region. This phenomenon does not affect the (true) ratio $J/\psi$ over
$DY$ (obtained in the NA50 standard analysis), but does affect the one
obtained by the so-called minimum bias analysis \cite{19r}. 

The results \cite{20r} of the model as a function of $E_{ZDC}$ are
presented in Fig.~2 (b). The centrality dependence of the $J/\psi$ 
suppression is reasonably well reproduced
except for the broad bump centered at $E_{ZDC} \sim 12 $ TeV. No
such structure is seen in the $E_{T}$ distribution at the same $b$
($E_{T} \sim 85$ GeV). Note that the suppression beyond the knee 
($E_{ZDC} \ \lsim \ 5$ TeV) is well reproduced by the model.
This was to be expected since the tail of the  $E_{ZDC}$
distribution is not affected by the presence of the $J/\psi$
trigger. This provides a consistency check of our interpretation
of the suppression versus $E_{T}$ beyond the knee.
Note also that
the data of Fig.~2 (b) for $E_{ZDC} < 28$~TeV exhibit a $J/\psi$
suppression significantly steeper
than the one in the NA50 absorption model -- indicating that the anomalous
suppression is already present
in very peripheral collisions.
This confirms the trend already observed in the suppression versus $E_{T}$.

\begin{figure}
\begin{minipage}{40mm}
\centering\epsfig{file=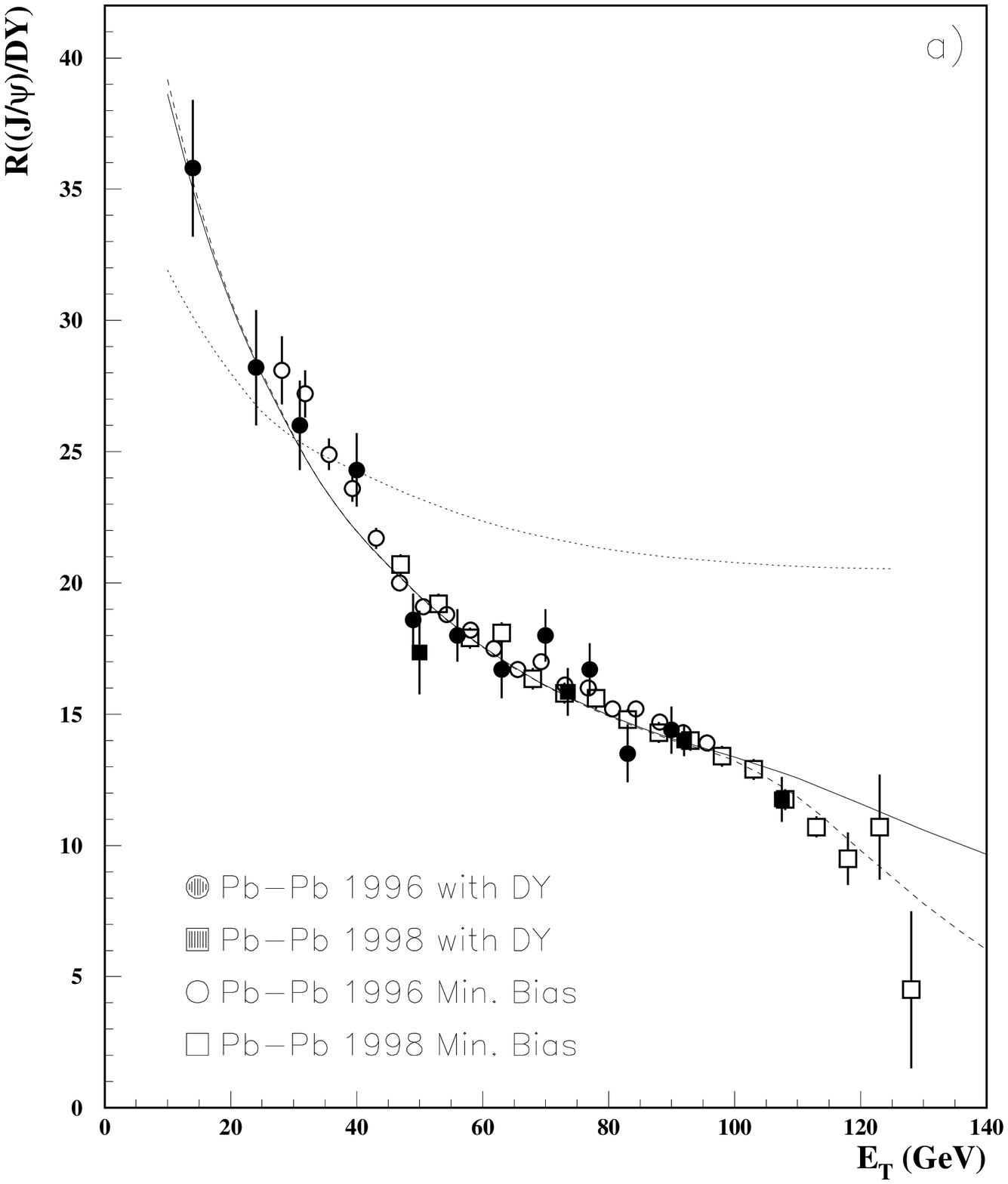,bbllx=0,bblly=30,bburx=540,bbury=620,height=3.0in}
\end{minipage}
\hspace{\fill}
\begin{minipage}{40mm}
\centering\epsfig{file=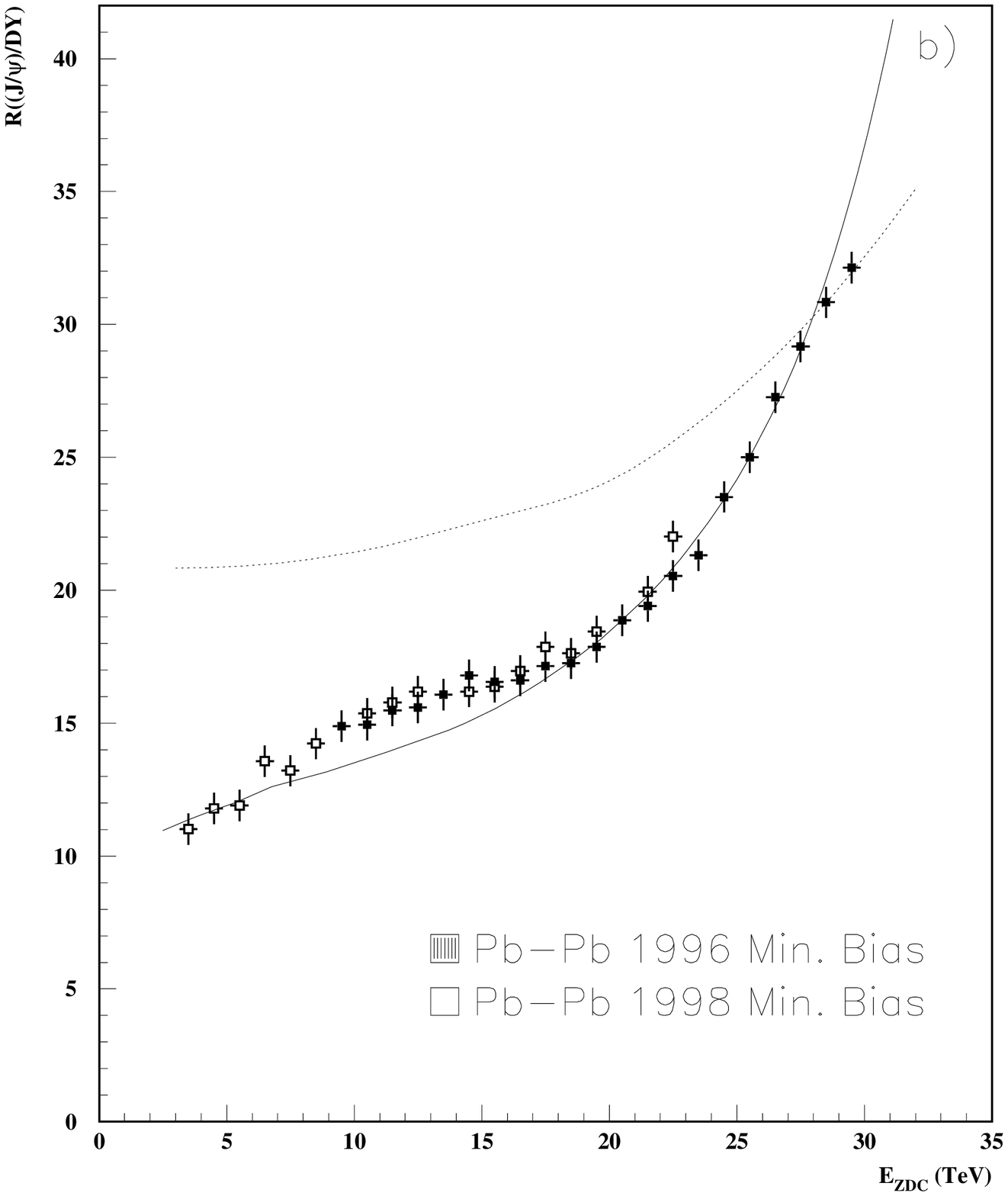,bbllx=280,bblly=30,bburx=840,bbury=620,height=3.0in}
\end{minipage}
\caption{
a): Ratio $J/\psi$ over DY versus $E_T$ compared to NA50 data
\protect\cite{17br}. The full curve is 
the theoretical prediction with
$E_{T}$ fluctuations and the dashed line contains $E_{T}$ fluctuations
and the $E_{T}$ loss induced by the 
$J/\psi$ trigger\protect\cite{19r}.
b): Ratio $J/\psi$ over DY versus $E_{ZDC}$ compared to preliminary
results presented by NA50 \protect\cite{17br}. 
The full line is obtained \protect\cite{20r} computing for
each $b$, the value of $E_{ZDC}$ and the value of the ratio $R$,
taking into account fluctuations and 
changing the normalization by a factor 0.94.
In both figures the dotted line is the NA50 absorption model,
fitting pA and SU. The used
parameters are $\sigma_{abs}$= 4.5 mb, 
$\sigma_{co}$= 1 mb, $q$= 0.62,
$\alpha$= 0.076 and $N_{f}$= 1.15 fm$^{-2}$.} 
\end{figure}

\section*{Acknowledgments}
It is a pleasure to thank the organizers for a nice and 
stimulating meeting.
D.S. thanks
Fundaci\'on Barrie de la Maza for
financial support.


\section*{References}
\begin{thebibliography}{99}

\bibitem{1r} A. Capella and D. Sousa, nucl-th/0101023,
to be published in Phys. Lett. {\bf B}.

\bibitem{5r} A. Capella, A. Kaidalov, J. Tran Thanh Van, Heavy Ion
Physics {\bf 9} (1999).

\bibitem{10r} WA98 collaboration, M. M. Aggarwal et al, nucl-ex/0008004.

\bibitem{11r} PHENIX collaboration, K. Adkox et al, nucl-exp/0012008.

\bibitem{13r} P. Aurenche, F. Bopp, A. Capella, J. Kwiecinski, M.
Maire, J. Ranft and J. Tran Thanh Van, Phys. Rev. {\bf D45}, 92 (1992).

\bibitem{14r} N. Armesto and A. Capella, Phys. Lett. B430 (1998) 23. \\
N. Armesto, A. Capella and E. G. Ferreiro, Phys. Rev. C59 (1999) 359.

\bibitem{16r} A. Capella, E. G. Ferreiro and A. Kaidalov, Phys. Rev. Lett. 85 (2000) 2080.

\bibitem{17r} J.-P. Blaizot, P. M. Dinh and J. Y. Ollitraut, Phys. Rev. Lett. 85 (2000) 4020.

\bibitem{17br} NA50 collaboration, M. C. Abreu et al, Phys. Lett. B477 (2000) 28;
{\it ibid.} Proceedings Quark Matter 2001 (presented by P. Bordalo).

\bibitem{19r} A. Capella, A. B. Kaidalov and D. Sousa,
nucl-th/0105021, submitted to Phys. Rev. C.

\bibitem{20r} A. Capella and D. Sousa, in preparation.

\end{thebibliography}
\end{document}